\documentclass[]{llncs}
\usepackage{graphicx}
\pdfoutput=1
\usepackage[utf8x]{inputenc}
\usepackage{epstopdf}
\usepackage{algorithm,algorithmic}
\usepackage{complexity}
\usepackage{amsmath,amssymb,amsfonts}
\usepackage{textcomp}
\usepackage{xcolor}
\begin{document}
\title{ Results on Competitiveness of Online Shortest Remaining Processing Time(SRPT) Scheduling with Special Classes of Inputs }
%
%
\author{Sheetal Swain \and
Rakesh Mohanty \and Debasis Dwibedy}
\authorrunning{S. Swain, R.Mohanty and D. Dwibedy}
%
\institute{Veer Surendra Sai University of Technology, Burla, 768018, Odisha, India
\email{\{sheetalswain18, rakesh.iitmphd, debasis.dwibedy\}@gmail.com}}
%
\maketitle              
\begin{abstract}
Shortest Remaining Processing Time (SRPT) is a well known preemptive scheduling algorithm for uniprocessor and multiprocessor systems. SRPT finds applications in the emerging areas such as scheduling of client's requests that are submitted to a web server for accessing static web pages, managing the access requests to files in multiuser database systems and routing of packets across several links as per bandwidth availability in data communications. SRPT has been proved to be optimal for the settings, where the objective is to minimize the mean response time of a list of jobs. According to our knowledge, there is less attention on the study of online SRPT with respect to the minimization of makespan as a performance criterion. In this paper, we study the SRPT algorithm for online scheduling in multiprocessor systems with makespan minimization as an objective.  We obtain improved constant competitiveness results for algorithm SRPT for special classes of online job sequences based on practical real life applications. 
\end{abstract}%
\section{Introduction}\label{sec:Introduction}
The inherent nature of input availability in current practical applications has paved the way for study of online algorithms [1]. In real world applications, inputs are given one at a time and an online algorithm  processes the available input irrevocably prior to the availability of the next input. An  online algorithm has no knowledge of future inputs and makes decision based on the current and past inputs. A sequence of inputs are processed on the fly to produce the final output. Hence, the performance of an online algorithm depends on the order of the input sequence rather than the size of the input sequence. Therefore, design and analysis of efficient online algorithms have been prompted as a non-trivial research challenge. \\ 
\textit{Competitive Analysis} [2] has been considered as the standard and most widely used performance measure for online algorithms. In competitive analysis, the cost obtained by an online algorithm is compared with that of the cost incurred by the optimum offline algorithm [2]. An optimum offline algorithm always incurs minimum cost among all the offline algorithms. The maximum ratio between the cost of an online algorithm and the cost of the optimum offline algorithm is defined as the \textit{competitive ratio}. The competitive ratio is valid for all permutations of online job sequences. For minimization problems, the \textit{lower value} ($\geq 1$) of the competitive ratio implies improved online algorithm. Efficient online algorithms have  competitive ratios bounded by a constant. Our objective is to obtain optimal competitive ratio, which is bounded by a constant.\\
The classical problem of scheduling is of great practical and research interest. In case of uniprocessor scheduling, there is only one processor and multiple jobs. Here, the scheduler has to decide which job to schedule next on the processor, depending upon certain parameters and constraints. In multiprocessor scheduling, we have multiple processors and multiple jobs. The scheduler has to decide which job to schedule next on which processor [1]. 
In case of offline scheduling, the arrival time of all jobs is assumed to be zero i.e., all jobs are available before the execution of the algorithm. Here, the scheduler has prior information regarding the whole input sequence of jobs. In contrast, in case of online scheduling, the jobs arrive over time, and at any instance of time, the scheduler has the knowledge of current and past jobs only [3,4], the future jobs being unknown. Online Scheduling finds applications in a wide range of real life scenarios such as university timetabling, client-server systems,  airline and train scheduling etc.\\  
The problem of multiprocessor scheduling has been proved to be NP-Hard [5]. Many researchers have attempted to find near optimal solutions for online scheduling  [6-7, 9-12]. To understand the variants of online scheduling and overview of important results, one can go through the recent survey article in [8].\\  In this paper, we  study the Shortest Remaining Processing Time (SRPT) [13] online scheduling algorithm in the multiprocessor setting and perform the competitive analysis based on some non-trivial online job sequences.
\section{Background and SRPT Algorithm}\label{sec:Background and SRPT Algorithm}
\subsection{Basic Terminologies and Notatios}\label{sec:Basic Terminologies and Notatios}
We present basic terminologies and notations as follows.
\begin{itemize}
\item $J_i, 1 \leq i \leq n$, denote $i^{th}$ job, where $n$ is the total number of jobs. Here jobs are independent to each other.
\item $P_j, 1 \leq j \leq m$, denote the $j^{th}$ processor, where $m$ is the number of processors. All processors are identical (i.e., processing speed of all processors is equal).
\item $T_i, 1 \leq i \leq n$, denote the processing time of $J_i$.
\item Makespan ($w$), denote the completion time of the job that finishes last in the schedule.
\item $w_{SRPT}$ is the makespan incurred by algorithm SRPT.
\item  $w_{OPT}$ is the makespan obtained by optimal offline algorithm $OPT$.
\item 	${CR}_{SRPT} = \frac{w_{SRPT}}{w_{OPT}}$, denote the competitive ratio of SRPT.
\end{itemize}
\subsection{SRPT Algorithm}\label{sec:SRPT Algorithm}
Shortest Remaining Processing Time (SRPT) is a preemptive scheduling algorithm, in which the job having the least remaining processing time is executed first. Here, we are given $m$ identical processors and a set of $n$ independent jobs with equal processing time ( each with t units ), where each job is given at an interval of $1$ unit each. The job which arrives first is executed on the processor $P_1$. If multiple jobs arrive simultaneously, then the job having smallest processing time is scheduled on the processor having smaller index among all the idle processors. Once a processor $P_i$ starts executing a job, it is interrupted only when a new job arrives. The processor $P_i$ then scans the list of all available jobs in search of the job with the shortest remaining processing time and starts its execution (or resumes execution of the current job). The job with shorter remaining processing time is always assigned to the idle processor having smaller index. All the processors start scanning for jobs at time $t=0$, and proceed in the manner mentioned above, until the execution of all jobs is completed.
Furthermore, we consider the following constraints for execution of SRPT algorithm.
\begin{itemize}
\item
Number of jobs is greater than or equal to the number of processors. ($n\geq m$)
\item Processing time of each job is greater than or equal to the number of processors. ($t\geq m$)
\end{itemize}
\section{Our Competitive Analysis Results for Algorithm SRPT based on Special Classes of Input Job Sequences}
\label{subsec:Our Competitive Analysis Results for Algorithm SRPT based on Special Classes of Input Job Sequences}
\begin{figure}[h]
\centering
\includegraphics[scale=1.3]{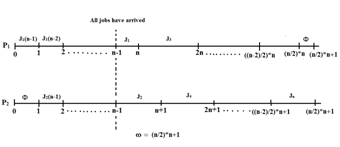}
\caption{Timing Diagram of SRPT for class 2 with even number of Jobs}
\label{fig:ganttchart3.png}  
\end{figure}

\begin{figure}[h]
\centering
\includegraphics[scale=1.3]{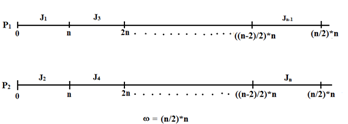}
\caption{Timing Diagram of of OPT for class 1 with Even number of Jobs }
\label{fig:ganttchart4.png}  
\end{figure}

\begin{figure}[!ht]
\centering
\includegraphics[scale=1.0]{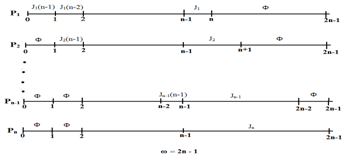}
\caption{Timing Diagram of SRPT for class 1 with $m=n$}
\label{fig:ganttchart5.png}  
\end{figure}

\begin{figure}[h]
\centering
\includegraphics[scale=1.0]{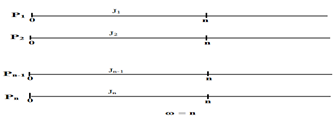}
\caption{Timing Diagram of OPT for class 2 with $m=n$}
\label{fig:ganttchart6.png}  
\end{figure}

\begin{figure}[h]
\centering
\includegraphics[scale=1.1]{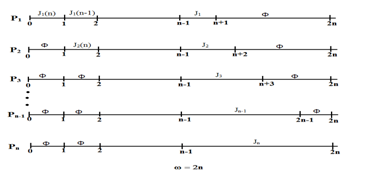}
\caption{Timing Diagram of SRPT for class 2}
\label{ganttchart7.png}  
\end{figure}

\begin{figure}[h]
\centering
\includegraphics[scale=1.2]{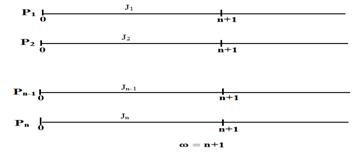}
\caption{Timing Diagram of OPT for class 2}
\label{ganttchart8.png}  
\end{figure}

\begin{figure}[h]
\centering
\includegraphics[scale=1.0]{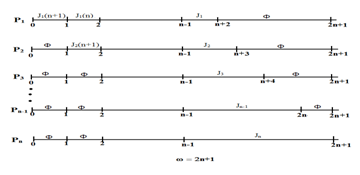}
\caption{Timing Diagram of SRPT for class 3}
\label{ganttchart9.png}  
\end{figure}

\begin{figure}[h]
\centering
\includegraphics[scale=1.0]{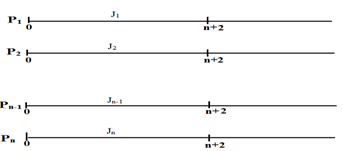}
\caption{Timing Diagram of OPT for class 3}
\label{ganttchart10.png}  
\end{figure}

\begin{figure}[!htbp]
\centering
\includegraphics[scale=0.8]{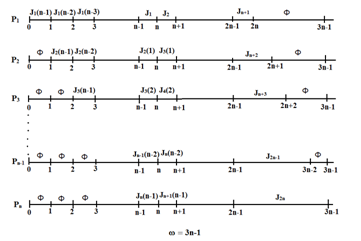}
\caption{Timing Diagram of SRPT for class 4}
\label{ganttchart17.png}  
\end{figure}

\begin{figure}[!th]
\centering
\includegraphics[scale=1.0]{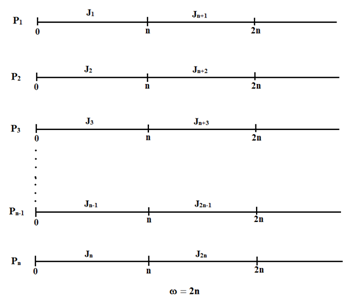}
\caption{Timing Diagram of OPT for class 4}
\label{ganttchart18.png}  
\end{figure}

We analyze the performance of algorithm SRPT by considering special classes of online job sequences. We consider that a sequence of jobs arrive one by one at an unit time interval. The special classes of input job sequences are judiciously chosen and presented as follows.\\
\textbf{Class $S_1$:} The online job sequence consists of $n$ jobs, where each job has its processing time of $n$ unit.\\
\textbf{Class $S_2$:} The online job sequence consists of $n$ jobs, where the processing time of each job is $n+1$ unit.\\
\textbf{Class $S_3$:} The online job sequence consists of $n$ jobs, where the processing time of each job is $2n$ unit.\\
\textbf{Class $S_4$:} The online job sequence consists of $2n$ jobs, where each job has its processing time of $n$ unit.\\
\textbf{Class $S_5$:} Here, we consider a scenario where the processing time of the jobs is equal to the number of jobs. For instance, we have $2n$ jobs and $n$ processors, where each job has the processing time of $2n$ unit.\\\\ 
\textbf{Theorem. 3.1.} \textit{Algorithm SRPT is   $1.5$-competitive for $S_1$  with $m=2$, where $n$ is an even integer.}\\\\
\textit{Proof:} Let $w_{SRPT}$ and $w_{OPT}$ be the makespans obtained by algorithms SRPT  and OPT respectively. \\
\textit{Computation of $w_{SRPT}$:} Suppose, the job $J_1$ arrives at time $t=0$, and is scheduled on the processor $P_1$. Subsequently, job $J_2$ arrives at time $t=1$. In this interval of time, processor $P_2$ is idle. When $J_2$ arrives, both the processors start scanning the list of available jobs for the shortest remaining processing time. Since $J_1$ has the shortest remaining processing time, it is scheduled on the smaller index processor, ( i.e., $P_1$ ) and $J_2$ is scheduled on processor $P_2$. Similarly, when job $J_3$ arrives at time $t=2$, the processors again scan the list of available jobs, and this continues till time $t=n-1$, ( i.e., till availability of all jobs ). After time $t> n-1$, each scheduled job is executed on the processor, only after the execution of the previously scheduled job. Since the number of jobs is even, the makespan incurred by algorithm SRPT is $w_{SRPT}= \frac{n}{2}(n+1)$. We present the execution schedule for $S_1$, constructed by algorithm SRPT through a timing diagram in Figure \ref{fig:ganttchart3.png}.
\\
\textit{Computation of $w_{OPT}$:} In the offline scenario, the arrival time of all the jobs is assumed to be zero, and the jobs are scheduled on the processors according to their processing times. In this case, since the processing time of all  jobs is same, so they are scheduled on the processors according to their index numbers. $J_1$, having the smallest index number is scheduled on the smaller index processor ( i.e., $P_1$ ) and so on. Each scheduled job is executed, only after the execution of the previously scheduled job. Since the number of jobs is even, the makespan incurred by optimal offline algorithm is $w_{OPT} = \frac{n}{2}(n)$. The optimum execution schedule for $S_1$ is shown by a timing diagram in Figure \ref{fig:ganttchart4.png}.
\\
$CR_{SRPT}=\frac{w_{SRPT}}{w_{OPT}}=\frac{\frac{n}{2}(n+1)}{\frac{n}{2}(n)}= \frac{n^2+2}{n^2}=\frac{3}{2}-\frac{n^2-4}{2n^2} \leq 1.5$. \hfill\(\Box\)\\\\
\textbf{Theorem 3.2.} \textit{Algorithm SRPT is ($2-\frac{1}{n}$)-competitive for $S_1$, where $m=n$.}\\\\
\textit{Proof.} 
\textit{Computation of $w_{SRPT}$.} Job $J_1$ is given at time $t=0$ and is scheduled on processor $P_1$. Job $J_2$ is given at time $t=1$. In the time interval from 0 to 1, processor $P_1$ executes job $J_1$ and at this time all other processors from $P_2$ to $P_n$ are in idle state, since there is no other job to be executed. When job $J_2$ is given, $P_1$ halts the execution of $J_1$ and scans the list of available jobs for the shortest remaining time. All other processors also scan the list of available jobs. Since, $J_1$ has the shortest remaining processing time, i.e., $n-1$ unit, it is scheduled on the smaller index processor, i.e, on $P_1$. Similarly, job $J_2$ is scheduled on processor $P_2$, and all other processors are idle until any other job arrives. We continue scheduling in the same fashion til all jobs arrive, i.e. up to time $t=n-1$. After time $t=n-1$, each job is scheduled on the processor, only after the completion of the previously scheduled job. From time $t=n$ till the final completion time, $P_1$ remains idle, since there is no other job to be scheduled. Similar scenario happens in case of other processors also. Therefore, SRPT obtains a makespan of $w_{SRPT}$ = $2n-1$. We present the schedule construction of SRPT in Figure \ref{fig:ganttchart5.png}.\\
\\
\textit{Computation of $w_{OPT}$:} All jobs are given at time $t=0$. Thus, according to the principle of SRPT algorithm, the job with the shortest remaining processing time must be scheduled on the lower index processor. But in this case, all  jobs have equal processing time. Therefore, jobs are scheduled on the processors according to their index numbers. $J_1$ is scheduled on $P_1$, $J_2$ is scheduled on $P_2$ and following this $J_n$ is scheduled on $P_n$. Therefore, the makespan incurred by OPT is: $w_{OPT}$ = $n$. We present the optimum schedule construction in Figure \ref{fig:ganttchart6.png}. \\ 
\\
We now have ${CR}_{SRPT} =  \frac{w_{SRPT}}{w_{OPT}}$ = $\frac{2n-1}{n}$= $2 - \frac{1}{n}$. \hfill\(\Box\)\\\\
\textbf{Theorem 3.3.} \textbf{Algorithm SRPT is ($2-\frac{2}{n+1}$)-competitive for $S_2$} \\\\
\textit{Proof.} 
\textit{Computation of $w_{SRPT}$:} Job $J_1$ is given at time $t=0$ and is scheduled on processor $P_1$. Job $J_2$ arrives at time $t=1$. In the time interval from 0 to 1, processor $P_1$ executes $J_1$ and all other processors from $P_2$ to $P_n$ are remains idle, since there is no other job to be scheduled. When $J_2$ arrives, $P_1$ halts the execution of $J_1$ and scans the list of available jobs for the shortest remaining time. All other processors also scan the list of available jobs. Since, $J_1$ has the shortest remaining processing time, that is, $n-1$ units, it is scheduled on the smaller index processor, that is, $P_1$. Similarly, $J_2$ is scheduled on processor $P_2$, and all other processors are idle till any other process arrives. This process is carried out till all the jobs have arrived, that is till time $t=n-1$. After time $t=n-1$, each job is scheduled on the processor, only after the execution of the previously scheduled job is completed. From time $t=n$ till the final completion time, $P_1$ remains idle, since there is no other job to be scheduled. Similar scenario happens in case of other processors also. Therefore, the makespan incurred by SRPT is: $w_{SRPT}$ = $2n$. We now sketch the schedule obtained by algorithm SRPT in Figure \ref{ganttchart7.png}.\\
\textit{Computation of $w_{OPT}$:} All the jobs have arrived at time $t=0$. Thus, according to the principle of SRPT algorithm, the job with the shortest remaining processing time must be scheduled on the lower index processor. But in this case, all the jobs have equal processing time, and thus the jobs are scheduled on the processors according to their index numbers. $J_1$ is scheduled on $P_1$, $J_2$ is scheduled on $P_2$, and so on. Therefore, the makespan obtained by OPT is $w_{OPT}$ = $n+1$. The schedule obtained by OPT for $S_2$ is presented in Figure \ref{ganttchart8.png} 
\\
We now have ${CR}_{SRPT} =  \frac{w_{SRPT}}{w_{OPT}}$ = $\frac{2n}{n+1}=2-\frac{2}{n+1}$.\hfill\(\Box\)\\\\
\textbf{Theorem 3.4.} \textit{SRPT is ($2-\frac{3}{n+2}$)-competitive for $S_3$}.\\\\
\textit{Proof Sketch.} Following the same pattern described in Theorem 3.1 to Theorem 3.3, we sketch the schedule for algorithm SRPT in Figure \ref{ganttchart9.png}. The timing diagram at Figure \ref{ganttchart9.png} shows that Algorithm SRPT achieves $w_{SRPT}= 2n+1$ for the input sequence $S_3$. 
\\
\textit{Computation of Optimum Offline:} We construct the schedule produced by OPT in Figure \ref{ganttchart10.png}. The makespan obtained by OPT i.e. $w_{OPT}= n+2$. \\
 \\
We now have ${CR}_{SRPT} =  \frac{w_{SRPT}}{w_{OPT}}$ = $\frac{2n+1}{n+2}=2-\frac{3}{n+2}$\\\\
\textbf{Theorem 3.5.} \textit{Algorithm SRPT is ($\frac{3}{2}-\frac{1}{2n}$)-competitive for $S_4$.}\\\\
\textit{Proof Sketch.} Computation of $w_{SRPT}$. The schedule achieved by algorithm SRPT for $S_4$ is shown in Figure \ref{ganttchart17.png}. In this setting, algorithm SRPT obtains the makespan $w_{SRPT} = 3n-1$. 
\\
\textit{Computation of $w_{OPT}$:} The schedule construction by OPT is shown in Figure \ref{ganttchart18.png}. Algorithm OPT incurs a makespan  $w_{OPT} = 2n$.
\\
We now have ${CR}_{SRPT} =  \frac{w_{SRPT}}{w_{OPT}}$ = $\frac{3n-1}{2n}=\frac{3}{2}-\frac{1}{2n}$. \hfill\(\Box\)
\section{Conclusion}\label{sec:Conclusion}
We analyzed the performance of  SRPT algorithm for preemptive online scheduling in multiprocessor systems by considering five special classes of inputs  in diversified frameworks based on the number of processors, number of jobs and processing times of the jobs.  We  obtained constant completitiveness results for SRPT algorithm  by considering makespan as an objective. More realistic characterization of inputs based on practical applications can be explored to derive stronger upper bound and lower bound results on the competitiveness of SRPT algorithm.
%
%
%

\begin{thebibliography}{}


\bibitem{A.Borodin}
A. Borodin, R. El-Yaniv (1998), Online Computation and Competitive Analysis. Cambridge University Press, Cambridge.
\bibitem{Robert}
R. E. Tarjan and S. S. Sleator (1985). Amortized Computational Complexity. SIAM Journal on Algebric and Discrete Methods, 6(2), pp.  306-318.
\bibitem{R.L Graham}
R.L. Graham (1966). Bounds for Certain Multiprocessor Anomalies. Bell System Technical Journal, 45, pp. 1563-1581.
\bibitem{R.L Graham}
R. L. Graham (1969), Bounds on Multiprocessing Timing Anomalies, SIAM Journal on Applied Mathematics, vol. 17, pp. 416 – 429.
\bibitem{M.R Garey}
M. R. Garey and D.S. Johnson (1979). Computers and Intractability: A Guide to the Theory of NP-Completeness. Freeman, $1^{st}$ Edition.
\bibitem{R.Fleischer}
R.Fleischer and M. Wahl (2000). Online Scheduling Revisited. Journal of Scheduling, 3, pp. 343-353.
\bibitem{Y. Bartal}
Y. Bartal, A. Fiat, H. Karloff and R. Vohra (1992). New Algorithms for an Ancient Scheduling Problem. In Proceedings of the $24^{th}$ ACM Symposium on the Theory of Computing (STOC), Victoria, Canada, pp. 51-58.
\bibitem{}
D. Dwibedy and R. Mohanty (2020). Online Scheduling with Makespan Minimization: State of the Art Results, Research Challenges and Open Problems. arXiv. 2001.04698 [cs.OS], January 2020.
\bibitem{}
S. Albers (2003), Online Algorithms: A Survey, Mathematical Programming, vol. 97, pp. 3-26.
\bibitem{G. Galambos}
G. Galambos and G.J. Woeginger (1993). An Online Scheduling Heuristic with Better Worst-case Ratio than Graham's List Scheduling. SIAM Journal of Computing, 22(2), pp. 349-355.
\bibitem{S. Albers}
S. Albers (1999). Better Bounds for Online Scheduling. SIAM Journal on Computing, 29, pp. 459-473. 
\bibitem{C. Phillips}
C. Phillips, C. Stein, and J. Wein (1995), Scheduling Jobs That Arrive Over Time. In Proceedings of the 4th Workshop on Algorithms and Data Structures (WADS’95), vol. 955 of Lecture Notes in Computer Science (LNCS), Springer Verlag, pp. 86 – 97.
\bibitem{C. Chung}
C. Chung, T. Nonner, A. Souza (2010), SRPT is 1.86- Competitive for Completion Time Scheduling, In Proceedings of SODA, pp. 1373 – 1388.
\end{thebibliography}
%

\end{document}